\documentclass[namedreferences]{solarphysics}
\usepackage[optionalrh, natbib]{spr-sola-addons}
\usepackage{graphicx}
\usepackage{url}

\newcommand{\aap}{    {\it Astron. Astrophys.}}
\newcommand{\apj}{    {\it Astrophys. J.}}
\newcommand{\apjl}{   {\it Astrophys. J. Lett.}}
 
\newcommand{\araa}{		{\it Ann. Rev. Astron. Astrophys.}}
\newcommand{\pasj}{   {\it Pub. Astron. Soc. Japan}}
\newcommand{\solphys}{{\it Solar Phys.}}
\newcommand{\sovast}{ {\it Soviet  Astron.}} 
\newcommand{\ssr}{    {\it Space Sci. Rev.}} 

\begin{document}
\begin{article}
\begin{opening}
\title{Synthetic radio views on simulated solar flux ropes}
\author{A.A.~\surname{Kuznetsov}$^{1}$\sep
        R.~\surname{Keppens}$^{2}$\sep
				C.~\surname{Xia}$^{2}$}
\runningtitle{Synthetic radio views on simulated solar flux ropes}
\runningauthor{Kuznetsov, Keppens, and Xia}        
\institute{$^{1}$ Institute of Solar-Terrestrial Physics, Irkutsk 664033, Russia
                  e-mail: \url{a_kuzn@iszf.irk.ru}\\
           $^{2}$ CmPA, Department of Mathematics, KU Leuven, BE-3001 Leuven, Belgium
					        e-mail: \url{Rony.Keppens@wis.kuleuven.be} \url{Chun.Xia@wis.kuleuven.be}}
\begin{abstract}
In this paper, we produce synthetic radio views on simulated flux ropes in the solar corona, where finite-beta magnetohydrodynamic (MHD) simulations serve to mimic the flux rope formation stages, as well as their stable endstates. These endstates represent twisted flux ropes where balancing Lorentz forces, gravity and pressure gradients determine the full thermodynamic variation throughout the flux rope. The obtained models are needed to quantify radiative transfer in radio bands, and allow us to contrast weak to strong magnetic field conditions. Field strengths of up to 100 G in the flux rope yield the radio views dominated by optically thin free-free emission. The forming flux rope shows clear morphological changes in its emission structure as it deforms from an arcade to a flux rope, both on disk and at the limb. For an active region filament channel with a field strength of up to 680 G in the flux rope, gyroresonance emission (from the third-fourth gyrolayers) can be detected and even dominates over free-free emission at the frequencies of up to 7 GHz. Finally, we also show synthetic views on a simulated filament embedded within a (weak-field) flux rope, resulting from an energetically consistent MHD simulation. For this filament, synthetic views at the limb show clear similarities with actual observations, and the transition from optically thick (below 10 GHz) to optically thin emission can be reproduced. On the disk, its dimension and temperature conditions are as yet not realistic enough to yield the observed radio brightness depressions.
\end{abstract} 
\end{opening}

\section{Introduction}
The solar corona is intricately structured magnetically, and hosts myriads of magnetic arcades, loops and twisted flux ropes. Flux ropes are vital ingredients in many dynamical models of solar eruptions, where their internal as well as external magnetic field variation can render them unstable to kink \citep{torok05}, torus \citep{kliem06}, or even tilt-kink \citep{keppens14} ideal MHD instability routes. Twisted flux ropes, and the concave upward parts of their field topology, are also the preferred sites to host solar filaments.

Solar filaments (also known as prominences when observed above the limb) are large masses of relatively cool ($\lesssim 10\,000$ K) and dense (about $10^9-10^{11}$ $\textrm{cm}^{-3}$) plasma suspended in the hot corona \citep{hir85, zir89, tan95, lab10, mac10, par14}. Although the filaments have been studied since the end of the XIX century, their formation and stability are not fully understood yet \citep[e.g.,][]{par14, pri14}. Many filaments end up with a final eruption; the filament eruptions are often associated with the coronal mass ejections \citep{mun79, gil00, sub01, gop03}, although it is not clear which of these phenomena is primary and which one is secondary. In addition, filament eruptions in active regions can play an important role in the initiation and development of, at least, some types of solar flares \citep{ura02, gre06, gre14}.

Solar flux ropes and the filaments/prominences that reside within them show distinctive emission at radio frequencies. In particular, the microwave emission bands (frequencies of about $1-30$ GHz) are well covered by the current and upcoming radio observing facilities. Radio views at these frequencies allow us to study the free-free emission and absorption properties in the solar flux ropes and filaments, or even identify gyrolayers in sufficiently strong magnetic field concentrations. Flux ropes, with or without embedded prominences, can exist in quiescent as well as in active solar regions, and can thus be characterized by very different thermodynamic and magnetic properties. In this paper, we use magnetohydrodynamic models of solar flux ropes to generate synthetic radio images at centimetric radio wavelengths.
Simulating the thermal bremsstrahlung and gyroresonance emissions requires the knowledge of density, temperature and magnetic field variation throughout the flux rope structures. In practice, this requires using flux rope models that go beyond simplified force-free, or zero plasma beta assumptions, which by construction ignore the thermodynamic variation altogether. Nevertheless, zero-beta modeling is able to provide insight into the dominant flux rope dynamics, e.g., when studying sympathetic eruptions due to removal of overlying stabilizing flux in multiple flux rope configurations \citep{torok11}, or in reproducing the slow rise and partial eruption phase of split filament systems using a double flux rope configuration \citep{kliem14}. The same zero beta paradigm has been used to study how magnetic field arcades can be deformed into helical flux ropes by means of flux cancellation on the solar surface \citep{Amari99, Mackay06, Lione02}.

In this work, we use a modern variant of these early studies, where either isothermal, finite-beta MHD, or full thermodynamically consistent MHD simulations provide density-temperature-magnetic field conditions that represent force-balanced configurations. A brief survey of current observational findings on the solar filaments/prominences and their filament channels as seen at radio wavelengths is presented in Section \ref{s-review}. The MHD models are discussed in Section \ref{s-mhd}, and the way to generate synthetic radio views from them is presented in Section \ref{s-radio}. The obtained radio views on the virtual flux ropes and filaments are presented in Sections \ref{res-isothermal} and \ref{res-filament}, respectively. The conclusions and outlook for future work are given in Section \ref{conclusion}.

\section{Microwave observations of filaments and flux ropes}\label{s-review}
\subsection{Filaments}
The filaments can be observed in a broad range of wavelengths. In the H$\alpha$ spectral line, they are seen as bright structures (prominences) above the limb or as dark filaments when observed against the solar disk. In the Extreme Ultraviolet (EUV) emission, the appearance of the filaments is wavelength-dependent as it reflects the formation temperatures of the EUV lines \citep{par14}. In radio observations, the filaments were firstly detected at millimetric and centimetric wavelengths \citep{kha64, kun70, kun72, dra70}, and later at lower and higher frequencies as well \citep{kun86, bas93, mar04, per13}. To date, the most comprehensive observations have been performed in the microwave range using the Nobeyama Radioheliograph and Siberian Solar Radio Telescope, with a particular focus on the eruptive processes \citep[etc.]{han94, gop97, gop98, han99, hor02, ura02, gop03, kun04, gre06, ali13, gop13}.

The dominant mechanism of generation/absorption of the radio emission in filaments is thermal bremsstrahlung (free-free emission and absorption); in addition, dense cores of the filaments block propagation of radio emission at the frequencies below the local plasma frequency. As a result, the appearance of the filaments in the microwave range (at the frequencies of ${\gtrsim 1}$ GHz) is usually similar to that in H$\alpha$: they look like dark structures (brightness depressions) when observed on the solar disk and like bright prominences when observed above the limb. 

At the same time, the quantitative characteristics of the microwave emission from the filaments depend on many factors and, first of all, on the considered emission frequency. At the frequencies of about $1-30$ GHz the filaments are expected to be optically thick, i.e., their observed brightness temperatures $T_{\mathrm{b}}$ should be nearly equal to their kinetic temperatures $T$. This has been confirmed by observations: the typical observed brightness temperatures of the filaments in the mentioned frequency range are of order of $5\,000-15\,000$ K both for the filaments on the disk and above the limb \citep[etc.]{but75, rao77, kun78, rao79, kun85, kun86, han94, gop97, gop98, ura02, gop03, kun04, gre06, gol08, ali13}. When multi-frequency observations of a particular filament are available, the brigthness temperature seems to be independent on the frequency. The filaments on the disk look dark because they absorb the emission from the underlying hotter layers of the solar atmosphere (the quiet Sun emission). Since the brightness temperature of the quiet Sun $T_{\mathrm{qS}}$ decreases with frequency, the on-disk filaments at lower frequencies, as a rule, demonstrate a higher contrast \citep{gre06}. Another important factor is the filament temperature: e.g., at the frequency of 17 GHz the filaments on the solar disk become invisible when heated to ${\sim 10\,000}$ K (the quiet Sun temperature at this frequency), although they remain visible (and even become brighter) above the limb \citep{han99, gop13}.

At higher frequencies, the free-free emission/absorption mechanism becomes less effective and at the frequencies of ${\gtrsim 30}$ GHz the filaments can become optically thin. As a result, we obtain $T_{\mathrm{b}}<T$ for the filaments above the limb and $T<T_{\mathrm{b}}<T_{\mathrm{qS}}$ (partial absorption of the quiet Sun emission) for the filaments on the solar disk. In this frequency range, the observed brightness temperatures of the filaments above the limb vary from the typical filament temperatures of $5\,000-10\,000$ K to much lower values. Similarly, the observed filaments on the disk vary in appearance from well-defined dark stripes to barely noticeable brightness depressions (relative to the quiet Sun level). Both the brightness temperature of the filaments above the limb and the relative brightness depression of the filaments on the disk tend to decrease with the frequency \citep[see][etc.]{kun70, kun72, but75, kun78, rao79, iri95, kun04, gre06}. The frequency dependence of the observed brightness temperature can be used to diagnose the temperature and density of the filaments \citep[e.g.,][]{iri95, chi01, gre06}.

Since the angular resolution in the radio range is in most cases insufficient to resolve directly the filament widths, these widths are estimated using deconvolution or forward-fitting when the observations are compared with the model images convolved with the instrument beam. The filament widths inferred from the microwave observations are variable: sometimes they are nearly the same as the widths observed in H$\alpha$ \citep[etc.]{but75, rao79, sch81, han94}, while sometimes they exceed the widths of the H$\alpha$ filaments by a factor of about $1-5$ \citep[etc.]{kun72a, chi75, rao77, kun78, kun86, iri95}; the observed widths of the radio filaments tend to decrease with the frequency. Most likely, this is caused by inhomogeneous structure of the filaments, since the radio emission can be affected not only by the dense and cool filament core, but also by the transition region between the filament and the surrounding corona, where the plasma density gradually decreases and the temperature increases \citep[e.g.,][]{kun78}; this transition in different events can be more or less sharp.

\subsection{Neutral-line associated sources}
The previous Section summarizes the microwave observations of well developed filaments with a cool dense core; the low temperature is the key factor affecting appearance of these filaments in the observations. At the same time, the hot overdense flux ropes (i.e., with the same temperature as the surrounding coronal plasma) are also likely to exist. For example, in the study of \citet{xiaiso14} evolution of a helical magnetic flux rope simulated in isothermal MHD model have led to concentration of plasma and magnetic field within the flux rope. Such simulations confirm that overdense flux ropes can be formed even in completely isothermal models, i.e., the isothermal approximation is well justified during most stages of the flux rope evolution; a nearly constant temperature is maintained due to the fast thermal conduction in the lower solar corona and slow change of magnetic field there. On the other hand, formation of a cool filament requires development of the thermal instability; once the threshold for this instability is reached (which can take a long time), the filament cooling happens very quickly \citep{xia14, kan15}. Therefore we conclude that the isothermal (or nearly isothermal) flux ropes with enhanced plasma density and magnetic field strength may be a common phenomenon in the solar corona.

The forming filament channels without filaments inside can be difficult to identify in observations. Since all filaments are located above the neutral lines of the photospheric magnetic field (although not all neutral lines give rise to filaments), we can expect some specific features associated with the neutral lines. Indeed, the so-called neutral-line-associated microwave sources (NLS) are a common phenomenon \citep[see, e.g., the papers of][and references therein]{ura08, yas14, bak15}. As the name implies, NLS occur in the vicinity of magnetic neutral lines in active regions; they are bright, compact and relatively stationary (can live up to several days). The NLS have typical brightness temperatures of a few MK (up to $\sim 10$ MK). The spectral observations reveal a sharp spectrum with the peak at $\sim 5-7$ GHz and a steep decrease at higher frequencies \citep{akh86, ali93, bog12, abramax15}; nevertheless, the brightness temperature can reach $\sim 1-2$ MK at the frequencies of $\sim 17-34$ GHz as well \citep{ura00, ura08, ura06, bak15}. Such spectra cannot be explained by the free-free emission mechanism; they are consistent with the thermal gyroresonance emission in a very hot plasma (and strong magnetic field) or with the nonthermal gyrosynchrotron emission.

The physical nature and formation mechanism of the NLS are yet unknown. According to the most popular explanation, the emission is produced in the arcade of magnetic loops overarching the filament channel \citep[][etc.]{kun81, str84, kor94}; this model is supported by observations of the NLS associated with the H$\alpha$ filaments \citep[][etc.]{kun77, kun80, kun81, ali82, str84, kun84, akh86, kor94}. On the other hand, the thermal free-free and/or gyroresonance emission from the flux rope (which is characterized by high plasma density and strong magnetic field) can make a significant or even dominant contribution. \citet{akh86, sych93, ura06} have concluded that the magnetic field strength in the NLS (estimated using the photospheric magnetograms and their force-free extrapolations) is insufficient to produce the gyroresonance emission at the observed GHz frequencies; the nonthermal gyrosynchrotron emission mechanism has been proposed (although it requires a stable long-living source of accelerated electrons). However, we note that the force-free approximation is essentially inapplicable to the flux ropes. As will be shown below, the MHD processes driven by the footpoint motions of a magnetic arcade can result in formation of strong magnetic fields in the corona. Obviously, making more definite conclusions about the nature of NLS and their relation to the filament channels requires observations with higher spatial resolution than now.

\section{Simulating radio views for flux ropes and filaments}
\subsection{MHD models}\label{s-mhd} 
In the remainder of this paper, we use three-dimensional magnetohydrodynamic (MHD) models of flux ropes, without and with actual filaments embedded, to create virtual radio views at varying radio frequencies. We should note that the MHD simulations are computationally expensive and take a long time; the choice of the simulation parameters is restricted by the technical considerations as well. Therefore we have studied only a limited number of MHD models, which illustrate different regimes of the flux rope formation. The model parameters have not been chosen to fit any particular observed event, but represent several general cases; therefore, a direct quantitative comparison of our results with observations would be difficult. Nevertheless, our results demonstrate qualitatively how the flux rope evolution affects the radio emission, and how the observable features are linked to the different source parameters.

The isothermal MHD models of helical magnetic flux ropes follow the approach presented in \citet{xiaiso14}, where they demonstrated the formation process of a large-scale flux rope in a quiet Sun region with finite plasma beta from an initially bipolar arcade. The initial arcade was extrapolated as linear force-free field from an artificial photospheric magnetogram. The deformation from a pure arcade to a flux rope configuration with overarching arcade was implemented by a combination of photospheric vortical and converging flow patterns enforced by the bottom boundary prescriptions. This ultimately led to a flux rope configuration with the magnetic field and plasma concentration within the flux rope.

Our first model (the ``low-field'' model corresponding to a quiet-Sun filament channel) is the same as in the study of \citet{xiaiso14}. In this model, the photospheric magnetogram had the extrema of $\pm 20$ G, the computational box dimensions were 240 Mm (length), 180 Mm (width), and 120 Mm (height), and the spatial grid resolution was 0.47 Mm. The simulations were performed until the maximum plasma velocity in the domain dropped below a certain threshold \citep[3.8 km $\textrm{s}^{-1}$, see][]{xiaiso14}, which was considered as the endstate. The evolution of the magnetic field and plasma in this model is shown in Fig. 4 in the paper of \citet{xiaiso14}. The resulting flux rope configuration is characterized by a low magnetic field strength typical for quiescent filament conditions: the vertical field component at the bottom of our domain, which is located at 3 Mm above the photosphere, attains values of up to $\pm 13$ G, while the field strength in the centre of the flux rope reaches 10 G. Together with the isothermal $T=1$ MK assumption and the plasma densities of $n=10^9$ $\textrm{cm}^{-3}$ and $1.8\times10^9$ $\textrm{cm}^{-3}$ at the bottom boundary and in the flux rope, respectively, this implies that free-free emission and absorption is the only relevant mechanism for the radio emission. Therefore the synthetic radio views in this case reflect only the plasma density structure from the MHD data. The isothermal MHD evolution self-consistently incorporates density changes and ultimately results in a delicate force-balance between Lorentz force, pressure gradient and gravity throughout the flux rope structure. This flux rope represents the magnetic structure of the filament channel of a quiescent filament. By taking snapshots corresponding to the times $t=27$, 53 and 80 (with the time unit chosen to be 86 seconds throughout this paper), we then obtain successive radio views on the forming flux ropes where the frames are about 38 minutes apart in real time; these radio views are presented in Section \ref{s-quiet}.

Besides the low-field case from \citet{xiaiso14} targeting quiescent filaments, we also analyse two more cases that represent smaller flux rope configurations at increased field strengths to mimic the magnetic structure of active region filaments. A medium field strength case has been generated by a procedure analogous to the low-field one, with magnetic field strength at the photosphere level reaching $\pm 100$ G, ultimately attaining about $60$ G in the flux rope centre. In this case, the bottom plane of the simulation box is located at 4 Mm above the photosphere representing the low corona where the field strength then reaches $\pm 54$ G; the (constant) temperature and density are similar to the low-field model. Finally, a high field strength case has been computed as well. In this case, the magnetic field strength reaches $\pm 1000$ G at the photosphere level, $\pm 520$ G at the bottom plane of the simulation box (in the low corona), and about 680 G in the flux rope; the density and temperature conditions are set to $n=2\times 10^9$ $\textrm{cm}^{-3}$ (at the bottom plane) and $T=1.5$ MK, respectively.

As said above, the simulations setup in the medium-field and high-field cases is similar to that in the study of \citet{xiaiso14}, but the computational box size is reduced to 120 Mm (length), 100 Mm (width), and 60 Mm (height), and the spatial resolution is reduced to 1 Mm. The reason is that the MHD simulations for an increased magnetic field strength (and the same spatial resolution) would require much shorter time steps to satisfy the stability criterion \citep{kep03, vdH07}. Due to the limited computational resources, we choose to reduce the resolution of the stronger field cases, although this increases the numerical resistivity and reduces the spatial accuracy. Ultimately, the main difference among the three models is in the 
attained maximum field strength, while the magnetic topologies (important for later filament formation) are very similar, despite of the differences in the box size, resolution, and thermodynamic conditions. Although the flux rope formation timescales differ in the considered three cases, here we are mainly concerned with the endstates obtained. In contrast to the low-field case \citep{xiaiso14}, the endstate in the medium- and high-field models is defined more qualitatively: as a state when the overall magnetic structure does not change anymore.

\begin{figure}
\centerline{\resizebox{10.2cm}{!}{\includegraphics{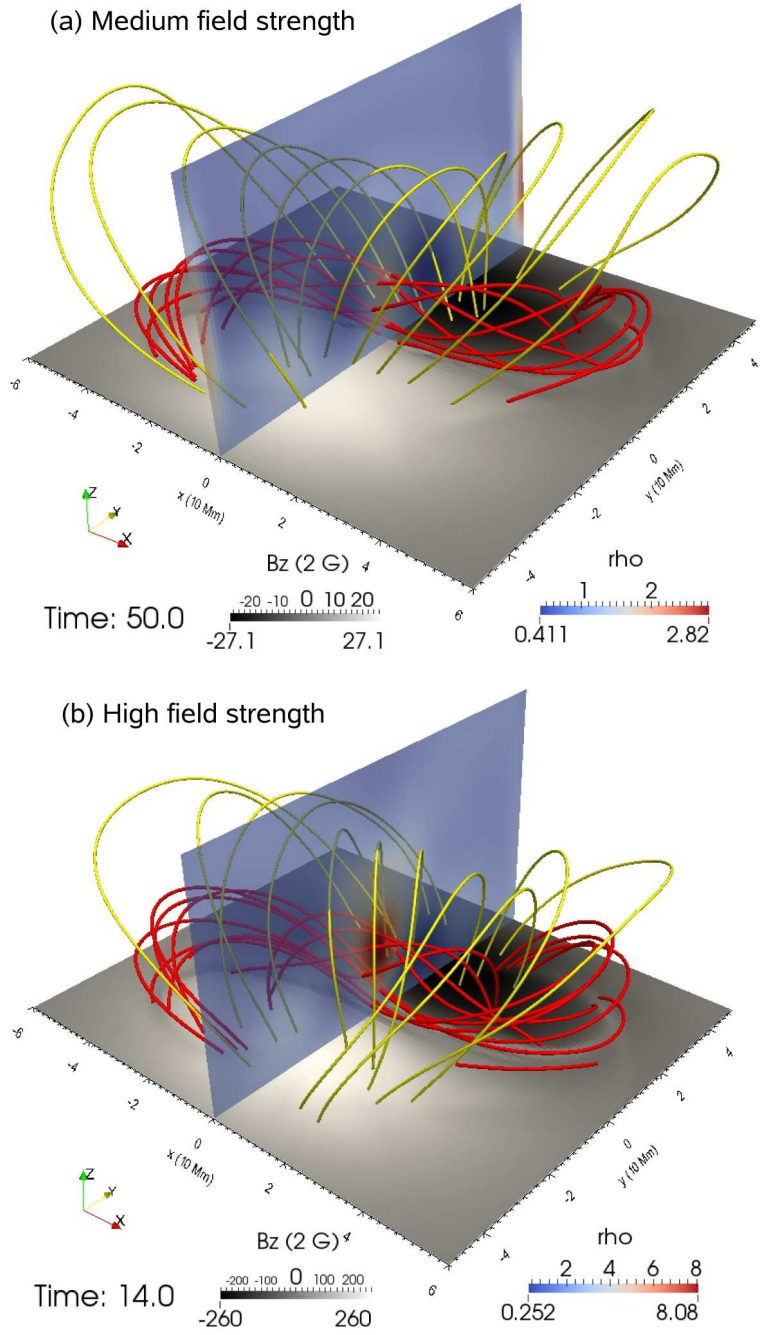}}}
\caption{Magnetic field lines of the medium (a) and the high (b) field
strength cases in their final endstates. In each panel, the red lines depict the flux rope and the yellow ones show the overlying arcade. The bottom horizontal plane of the simulation box is colored by vertical magnetic field strength and the central vertical plane shows the density distribution in units of $10^9$ $\textrm{cm}^{-3}$.}
\label{figchunMH}
\end{figure}

Figure \ref{figchunMH} gives an impression of both the medium- and high-field strength magnetic topologies and the corresponding plasma density distributions in their final stable flux rope endstates. In the adopted numerical time units these endstates are reached at $t=50$ and $t=14$ for the medium- and high-field cases, respectively. The MHD simulation results are used below to produce typical radio views for the flux ropes located either on the solar disk or at the limb; the synthetic radio views are presented in Sections \ref{ARMF} and \ref{ARHF} for the medium- and high-field models, respectively.

We should note that our models are based on the full set of MHD equations and thus go beyond the oft-used force-free approximation for the coronal magnetic field. This allows formation of strong magnetic fields in the corona; in particular, in the high-field model of an active region the magnetic field in the flux rope can become stronger than at the bottom of the simulation box. In turn, strong magnetic fields can result in formation of coronal gyroresonance emission sources.

\begin{figure}
\resizebox{12cm}{!}{\includegraphics{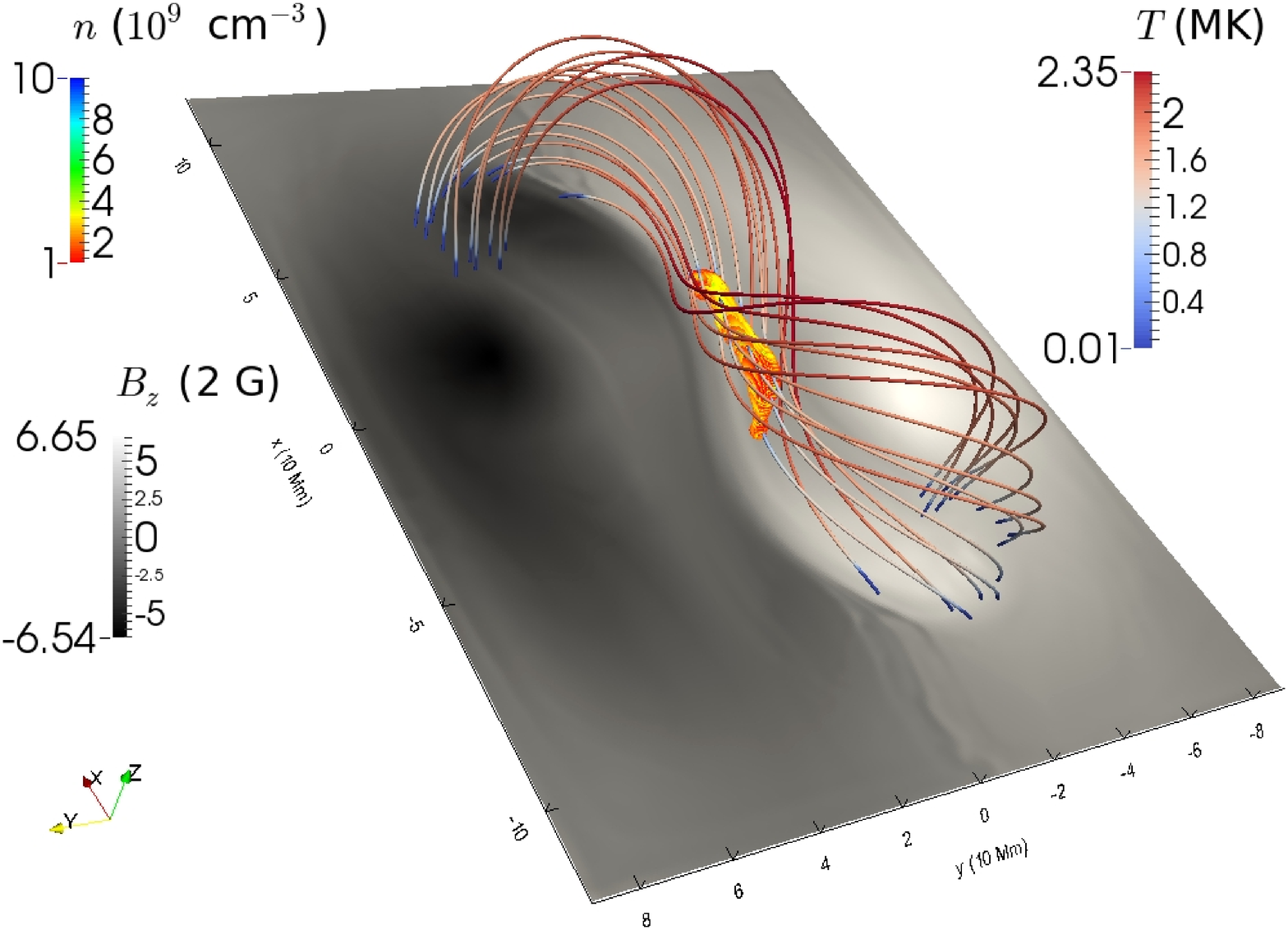}}
\caption{A filament embedded in a flux rope. The representative field lines are colored by temperature from blue to red. The filament is colored by density in a rainbow of colors and is presented by edges of grid cells where density exceeds $10^9$ $\textrm{cm}^{-3}$. The bottom plane is colored by vertical magnetic field in grayscale.}
\label{filamentxc}
\end{figure}

The final synthetic radio maps we present are based on the endstate of the fully thermodynamically consistent MHD study \citep{xia14}. In this case the low-field isothermal model was used as an input to study how optically thin radiative losses could trigger a full prominence formation within the dipped magnetic field region of the flux rope. A stable prominence structure with overall dimensions of $46.4\times 13.1\times 4.8$ Mm was established with the average density of $4.7 \times 10^9$ $\mathrm{cm}^{-3}$, i.e., well above the surrounding coronal density of $1.6\times 10^8$ $\mathrm{cm}^{-3}$. Although the structure is relatively small and low in mass, its temperature contrast with the corona is realistic with a $\sim 20\,000$ K filament embedded in a 1 MK corona. The magnetic topology and plasma distribution corresponding to the final state of the filament evolution are shown in Figure \ref{filamentxc}. In this work, this low-field model with embedded prominence is used to generate the first synthetic views on simulated filaments (see Section \ref{res-filament}).

\subsection{Emission mechanism}\label{s-radio}
As has been said above, the dominant mechanism determining the appearance of the solar filaments at radio wavelengths is the free-free emission and absorption. This process can be described by the radiation transfer equation
\begin{equation}\label{transfer}
\frac{\mathrm{d}I_f^{\sigma}}{\mathrm{d}l}=j_f^{\sigma}-\varkappa^{\sigma}I_f^{\sigma},
\end{equation}
where $I_f^{\sigma}$ is the spectral emission intensity of the wave mode $\sigma$ (either ordinary or extraordinary) and $l$ is the distance along the line of sight. The spectral emissivity $j_f^{\sigma}$ and the absorption coefficient $\varkappa^{\sigma}$ for the thermal bremsstrahlung process are given by the equations \citep[][etc.]{dul85, asc05, shi11, fle14}
\begin{equation}
\varkappa^{\sigma}=\frac{8e^6}{3\sqrt{2\pi}c(m_{\mathrm{e}}k_{\mathrm{B}})^{3/2}}\frac{n_{\mathrm{e}}^2\left<Z^2\right>\ln\Lambda}{N_{\sigma}f^2T^{3/2}},\qquad
j_f^{\sigma}=\varkappa^{\sigma}k_{\mathrm{B}}T\frac{N_{\sigma}^2f^2}{c^2},
\end{equation}
where $f$ is the emission frequency, $N_\sigma$ is the refraction index (under the considered conditions, it is close to unity for both wave modes), $T$ is the plasma temperature, $n_{\mathrm{e}}$ is the electron concentration, the factor $\left<Z^2\right>$ describes the average ion charge and $\ln\Lambda$ is the Coulomb logarithm. In the solar corona, the ion charge factor can be taken to be $\left<Z^2\right>=1.146$ at the temperatures above $35\,000$ K (the helium ionization temperature) or $\left<Z^2\right>=1$ below that temperature. The Coulomb logarithm (in the CGS units) equals approximately
\begin{equation}
\ln\Lambda\simeq\left\{\begin{array}{ll}
18.2+1.5\ln T-\ln f, & T<2\times 10^5~\textrm{K},\\
24.573+\ln T-\ln f, & T>2\times 10^5~\textrm{K}.
\end{array}\right.
\end{equation}
Note that the emission does not propagate (its absorption coefficient $\varkappa^{\sigma}$ goes to infinity) at the frequencies below the plasma frequency\footnote{More exactly, the cutoff frequency of the extraordinary mode slightly exceeds the plasma frequency. However, under the considered conditions this difference is negligible.}.

Another mechanism that can be responsible for the radio emission of solar filaments and filament channels is the gyroresonance emission of the thermal electrons gyrating in magnetic field \citep[e.g.,][]{zhe70, fle14}. In this case, the emission (and/or absorption) takes place at narrow resonance layers where the emission frequency coincides with a harmonic of the local electron cyclotron frequency $f_{\mathrm{B}}$ (i.e., $f=sf_{\mathrm{B}}$, $s=1, 2, 3,\ldots$). The change of the emission intensity at a given gyrolayer can be described by the equation
\begin{equation}\label{GR}
(I_f^{\sigma})_{\mathrm{out}}=(I_f^{\sigma})_{\mathrm{in}}\exp(-\tau^{\sigma}_s)+\frac{N_{\sigma}^2f^2}{c^2}k_{\mathrm{B}}T\left[1-\exp(-\tau^{\sigma}_s)\right],
\end{equation}
where $(I_f^{\sigma})_{\mathrm{in}}$ and $(I_f^{\sigma})_{\mathrm{out}}$ are the emission intensities before and after the gyrolayer, and $\tau^{\sigma}_s$ is the gyrolayer's optical depth:
\begin{equation}
\tau^{\sigma}_s\simeq\frac{\pi e^2n_{\mathrm{e}}}{fmc}\left(\frac{k_{\mathrm{B}}T}{mc^2}\right)^{s-1}\frac{s^{2s}N_{\sigma}^{2s-3}\sin^{2s-2}\theta}{2^{s-1}s!}\frac{(T_{\sigma}\cos\theta+L_{\sigma}\sin\theta+1)^2}{1+T_{\sigma}^2}L_{\mathrm{B}}.
\end{equation}
In the above equation, $\theta$ is the viewing angle (the angle between the local magnetic field and the line of sight), $T_{\sigma}$ and $L_{\sigma}$ are the components of the wave polarization vector \citep[see, e.g.,][]{fle10}, and $L_{\mathrm{B}}$ is the inhomogeneity scale of the magnetic field in the direction along the line of sight. Evidently, the gyroresonance emission requires a sufficiently strong magnetic field; in our simulations, it has been found to be important only in the high-field active region model (see Section \ref{ARHF}). Under the conditions typical of the solar corona, the gyrolayers corresponding to the harmonics with $s\le 3$ or $s>3$ are expected to be optically thick ($\tau_s^{\sigma}\gg 1$) or optically thin ($\tau_s^{\sigma}\ll 1$), respectively. As a result, the observed emission (with the brightness temperature equal to the local plasma temperature) is expected to be produced at the third gyrolayer, although under certain conditions other cyclotron harmonics can become dominant.

We calculate the two-dimensional emission maps by numerical integration of the radiation transfer equations (\ref{transfer}, \ref{GR}) along the selected lines of sight (corresponding to the image pixels). The initial intensity (at the far boundary of the simulation box) is set either to zero (for the filaments and flux ropes located at the limb) or to the typical emission intensity of the quiet Sun (for the sources on the solar disk); the average radio spectrum of the quiet Sun is taken from the paper of \citet{lan08}. The radio emission is calculated using the numerical code developed by \citet{fle14}, which takes into account the contributions of both the free-free and gyroresonance emission mechanisms. In addition to the above mentioned factors, this code automatically calculates the local tem\-pe\-ra\-tu\-re-dependent ionization degree (which affects the electron concentration $n_{\mathrm{e}}$) using the Saha equation \citep{saha20}; for the typical densities in the solar corona ($n\simeq 10^8-10^{11}$ $\textrm{cm}^{-3}$), the ionization degree rises steeply from nearly zero to nearly 100\% at the temperature of about $(5-7)\times 10^3$ K.

\section{Synthetic images}
As follows from observations and is supported by theory, the appearance of the filaments and filament channels in radio emission is frequency-dependent. On the other hand, the angular resolution of the radio telescopes is also frequency-dependent; in particular, in the decimetric/metric range both the existing instruments (e.g., the Nan\c{c}ay Radioheliograph) and the instruments expected soon will be unable to resolve any source structure for the considered MHD models. Therefore we consider here only the microwave emission (at the frequencies of about $3-30$ GHz). This frequency range is covered by such imaging instruments as the Siberian Solar Radio Telescope (5.7 GHz) and Nobeyama Radioheliograph (17 and 34 GHz); several instruments are currently under construction, such as the Expanded Owens Valley Solar Array ($1-18$ GHz), Upgraded Siberian Solar Radio Telescope ($3-24$ GHz) and Chinese Spectral Radioheliograph ($0.4-15$ GHz).

\begin{figure}
\includegraphics{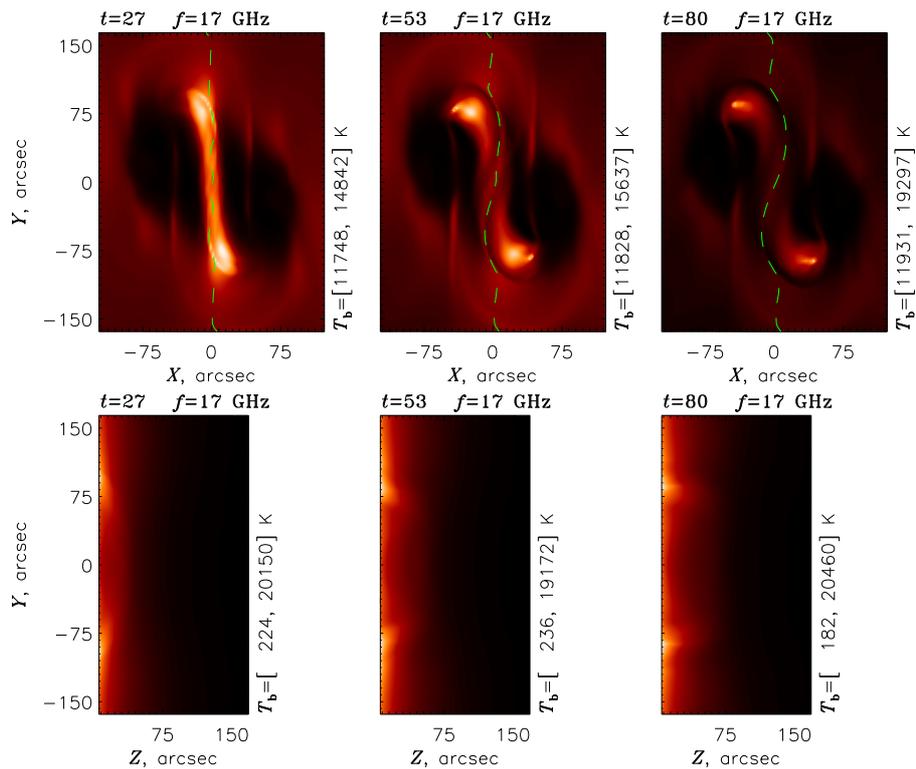}
\caption{Simulated radio images (at the frequency of 17 GHz) of the quiet-Sun filament channel at different times. Top row: the source region is located at the center of the solar disk; the dashed line represents the neutral line of the photospheric magnetic field. Bottom row: the source region is located at the limb. The brighthess temperature in each image varies from $T_{\mathrm{b}\,\min}$ (black) to $T_{\mathrm{b}\,\max}$ (white); the corresponding range of brightness temperatures $T_{\mathrm{b}}=[T_{\mathrm{b}\,\min}, T_{\mathrm{b}\,\max}]$ is given at the right side of each panel.}
\label{QuietEvol}
\end{figure}

\subsection{Emission from an isothermal filament channel}\label{res-isothermal}
\subsubsection{Quiet-Sun filament channel (low magnetic field)}\label{s-quiet}
Figure \ref{QuietEvol} demonstrates the simulation results for the quiet-Sun filament channel model. In this model, the magnetic field is weak (less than 20 G), the plasma density varies from $10^8$ to $3\times 10^9$ $\textrm{cm}^{-3}$; the plasma temperature is set to 1 MK. Three time moments illustrating different stages of the flux rope evolution are shown, in the simulation they are about 38 minutes apart. We have found that the free-free emission at the frequencies above 1 GHz is optically thin see the spectra in Figure \ref{SpectraFinalState}); the contribution of the gyroresonance emission is negligible. Therefore the computed images at all considered frequencies are nearly the same (except of different values of the brightness temperature) and coincide with the EUV images as well. In Figure \ref{QuietEvol}, the frequency of 17 GHz (one of the working frequencies of the Nobeyama Radioheliograph) is chosen.

If the filament channel is observed from above (i.e., located at the center of the solar disk) then at the beginning of the evolution process we see a bright stripe aligned along the photospheric neutral line and two darker areas to right and left from it. Later, the strongest emission gradually concentrates in two bright spots located at some distance from the neutral line; note an increase of the maximum radio brightness with time. However, the emission intensity is still relatively low: the relative radio brightening (after subtraction of the quiet Sun emission level) does not exceed 7500 K at 17 GHz and 90\,000 K at 5 GHz. Although such brightenings can be detectable, they are too weak in comparison with the observed neutral-line-associated sources in active regions.

If the source region is located at the solar limb then the images are dominated by two bright spots at the footpoints of the flux rope. The coronal emission is much weaker, which makes it hard to detect. Nevertheless, one can notice a faint arc-like structure that slowly ascends with time.

\begin{figure}
\includegraphics{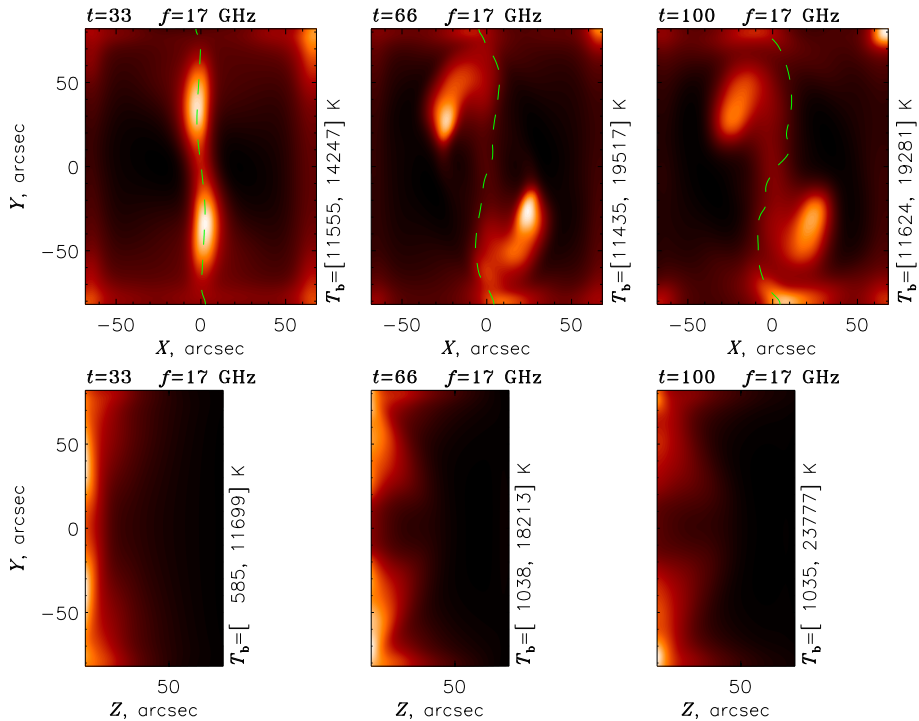}
\caption{Simulated radio images (at the frequency of 17 GHz) of the active-region filament channel (the medium-field model, with the magnetic field of up to 100 G) at different times. Top row: the source region is located at the center of the solar disk; the dashed line represents the neutral line of the photospheric magnetic field. Bottom row: the source region is located at the limb. At the right side of each panel, the range of brightness temperatures in the image is given.}
\label{ARMFevol}
\end{figure}

\subsubsection{Active-region filament channels: medium to high field cases}
\paragraph{Medium-field model}\label{ARMF}
Figure \ref{ARMFevol} demonstrates the evolution of an active-region filament channel. The magnetic field strength in the considered model can reach 100 G, which exceeds the typical values for the quiet Sun, but is still relatively weak for the solar active regions. The plasma density varies from $2\times 10^8$ to $5\times 10^9$ $\textrm{cm}^{-3}$; the plasma temperature is set to 1 MK. Like in the previous case, the free-free emission at the frequencies above 1 GHz is optically thin and the contribution of the gyroresonance emission is negligible; therefore we show only the 17 GHz images. The frames are taken at times $t=33$, 66 and 100, i.e., they also cover the deformation phase before the flux rope formation (the flux rope reaches a stable configuration at $t=50$, as shown in Figure \ref{figchunMH}, top panel).

Although the simulation box is now smaller than in the previous model, the source structure and evolution are qualitatively similar. Initially, the emission is produced mainly in a narrow stripe along the photospheric neutral line (when seen from above), and later concentrates in two bright spots corresponding to the footpoints of the forming flux rope. The brightness temperatures are also similar to the previous model, i.e., much lower than in the observed neutral-line-associated microwave sources. The limb view demonstrates a gradually rising arc-like structure.

\begin{figure}
\centerline{\includegraphics{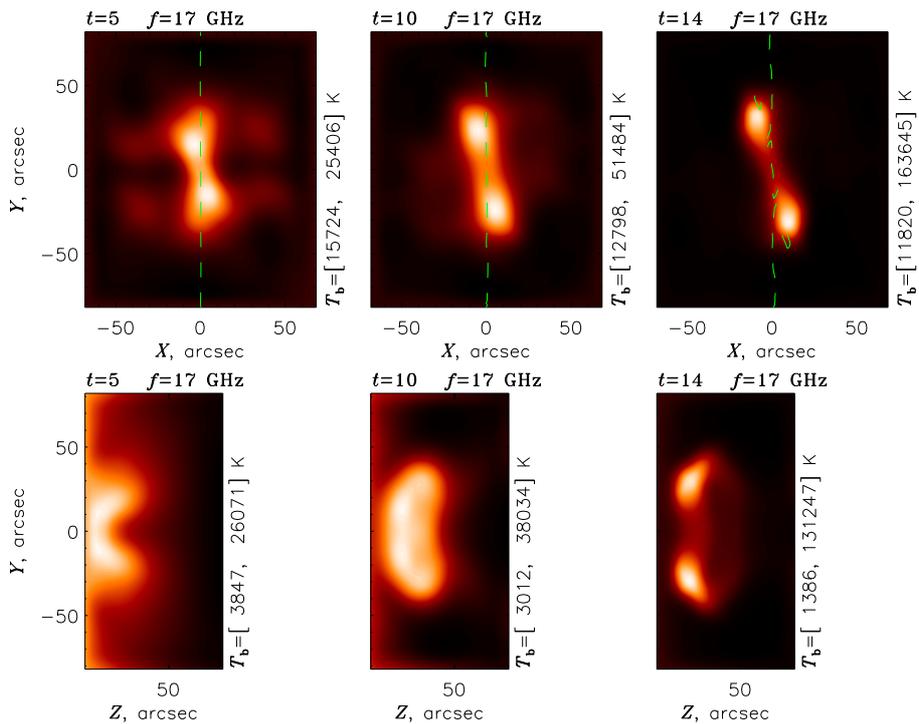}}
\caption{Simulated radio images (at the frequency of 17 GHz) of the active-region filament channel (the high-field model, with the magnetic field of up to 1000 G) at different times. Top row: the source region is located at the center of the solar disk; the dashed line represents the neutral line of the photospheric magnetic field. Bottom row: the source region is located at the limb. At the right side of each panel, the range of brightness temperatures in the image is given.}
\label{ARHFevol17}
\end{figure}

\begin{figure}
\centerline{\includegraphics{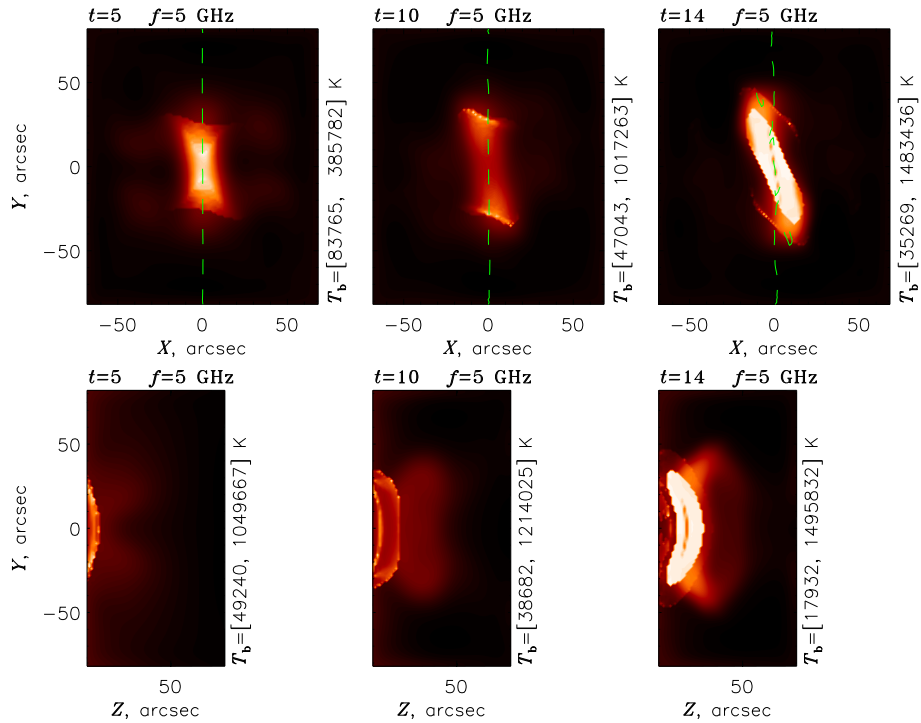}}
\caption{Same as in Figure \protect\ref{ARHFevol17}, for the emission frequency of 5 GHz.}
\label{ARHFevol5}
\end{figure}

\begin{figure}
\centerline{\includegraphics{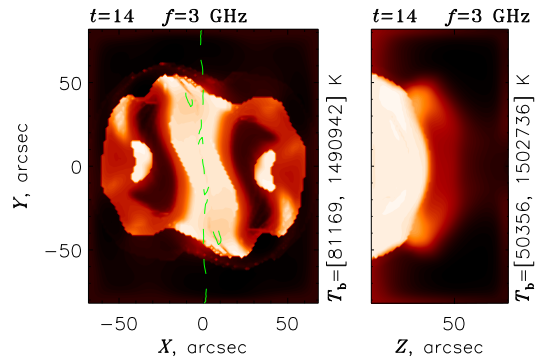}}
\caption{Simulated radio images of the active-region filament channel (the high-field model, same as in Figures \protect\ref{ARHFevol17}--\protect\ref{ARHFevol5}, at $t=14$) at the frequency of 3 GHz. Left: the source region is located at the center of the solar disk; right: the source region is located at the limb.}
\label{ARHFfinal3}
\end{figure}

\paragraph{High-field model}\label{ARHF}
Figures \ref{ARHFevol17}--\ref{ARHFfinal3} correspond to the high-field model of the active-region filament channel (Figure \ref{figchunMH}, bottom panel). Now, the magnetic field is stronger (up to 1000 G at the solar surface and up to 680 G within the flux rope). The plasma density varies from $6\times 10^7$ to $2\times 10^{10}$ $\textrm{cm}^{-3}$; the plasma temperature is set to 1.5 MK.

Figure \ref{ARHFevol17} demonstrates the evolution of the microwave emission at the frequency of 17 GHz, where the contribution of the gyroresonance emission is negligible (i.e., only the free-free emission is present). Like in the previous cases, the microwave images reflect the spatial distribution of the plasma within the simulated active region. The top view also reveals some similarity to the models with lower magnetic field: initially, the emission is concentrated in a short bright stripe aligned approximately along the photospheric neutral line; later, this stripe turns slowly with respect to the photospheric neutral line, and its ends gradually become brighter. In the limb view, we can see that the emission is produced mainly at some height above the solar surface; the evolution of the microwave images reflects the formation of an overdense region in the corona. As the plasma is accumulated within the flux rope, the emission brightness temperature gradually increases. This increase is much faster than in the previous (low-field and medium-field) cases, and the resulting brightness temperature is much higher as well: it can exceed 0.15 MK. Evidently, this is caused by the higher concentration of plasma within the flux rope (up to $2\times 10^{10}$ $\textrm{cm}^{-3}$). As a result, the free-free emission (although still optically thin, with $T_{\mathrm{b}}\ll T$) becomes much brighter. Our simulations indicate that, at least, some of the observed neutral-line associated sources with moderate brightness temperatures \citep[see, e.g.,][]{akh86, ura00, ura06, gre13} can be explained by the purely thermal free-free emission from overdense flux ropes.

The simulated images at higher frequencies (above 17 GHz) are not shown because they are nearly the same as in Figure \ref{ARHFevol17}, except of different values of the brightness temperature; this behaviour is typical of the optically thin free-free emission.

The contribution of the gyroresonance emission becomes noticeable (or even dominant) at the frequencies below $7-8$ GHz (see the spectra in Figure \ref{SpectraFinalState}), which affects the microwave images dramatically. Figure \ref{ARHFevol5} demonstrates the evolution of the microwave emission at the frequency of 5 GHz. Initially (at $t=5$ and 10), when the magnetic field in the core region of the forming flux rope is relatively weak, the emission at the considered frequency is produced mainly at the fourth gyrolayer. This gyrolayer is optically thin, i.e., the emission brightness temperature is lower than the plasma temperature; nevertheless, the contribution of the gyroresonance emission far exceeds that of the free-free emission. As a result, the microwave images reflect mainly the structure of the magnetic field in the active region. As the flux rope grows and its magnetic field strength increases, the third gyrolayer becomes the dominant emission source at the considered frequency; this gyrolayer is optically thick, i.e., the emission brightness temperature approaches the plasma temperature of 1.5 MK. At the final stage of the filament channel evolution (at $t=14$), the brightest (saturated) areas in the simulated microwave images represent the projections of the third gyrolayer, i.e., of the corresponding magnetic field strength isosurface ($B=595$ G for the emission frequency of 5 GHz). Note that, as can be seen in the limb view at $t=14$, this isosurface is located at some height in the corona and separated from the bottom of the simulation box. Such geometry reflects the above-mentioned fact that the considered model is not force-free; as a result, concentration of the magnetic field within the flux rope can create field strengths higher than in the lower layers. This effect enables generation of the neutral-line-associated sources due to the gyroresonance emission mechanism even in the active regions where the photospheric magnetic fields seem to be insufficient for that \citep[cf.][]{akh86, sych93, ura06}.

Figure \ref{ARHFfinal3} shows the simulated radio images at the frequency of 3 GHz (only the final state of the simulation at $t=14$ is shown). Again, the emission is produced mainly at the third gyrolayer. The brightest (saturated) areas in the images reflect the shape of this gyrolayer that corresponds now to the $B=357$ G isosurface; the brightness temperature reaches the plasma temperature of 1.5 MK.

The above simulations demonstrate that the gyroresonance emission from a hot dense flux rope with a strong magnetic field can be responsible, at least, for some of the observed neutral-line-associated microwave sources in the solar active regions. The brightness temperature of the emission in our simulations reaches 1.5 MK, which equals the assumed plasma temperature in the considered model. Such brightness temperatures are quite common for the observed neutral-line-associated sources, as reported, e.g., by \citet{kun80, kun81, kun84, akh86, ali92, ura08, bak15}. On the other hand, some observations have reported considerably higher brightness temperatures of the neutral-line-associated sources: up to $\sim 6-8$ MK \citep[][etc.]{ali82, akh86, sych93}. Evidently, such sources would require a higher plasma temperature, which seems to be feasible in the active regions \citep{yas14}.

As said above, the magnetic field strength in the considered model (in the flux rope) can reach 680 G. Therefore the gyroresonance emission mechanism is limited by the frequencies of up to $5.7-7.6$ GHz (for the emission at the third-fourth cyclotron harmonic). To produce emission at higher frequencies, we need a stronger magnetic field. For example, the emission at 17 GHz requires the field strength of $2024-1518$ G (again, for the emission at the third-fourth cyclotron harmonic), which is quite typical of the solar active regions. 

\subsection{Emission from a non-isothermal filament}\label{res-filament}
\begin{figure}
\includegraphics{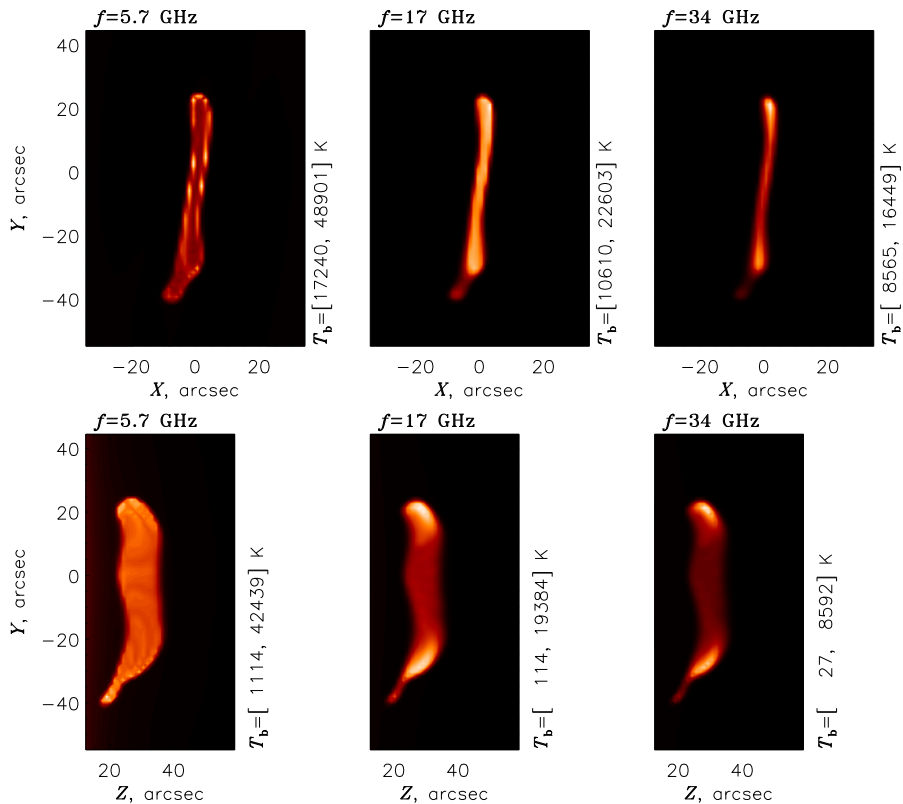}
\caption{Simulated radio images of the filament at three frequencies. Top row: the source region is located at the center of the solar disk. Bottom row: the source region is located at the limb. At the right side of each panel, the range of brightness temperatures in the image is given.}
\label{FilamentFinal}
\end{figure}

\begin{figure}
\includegraphics{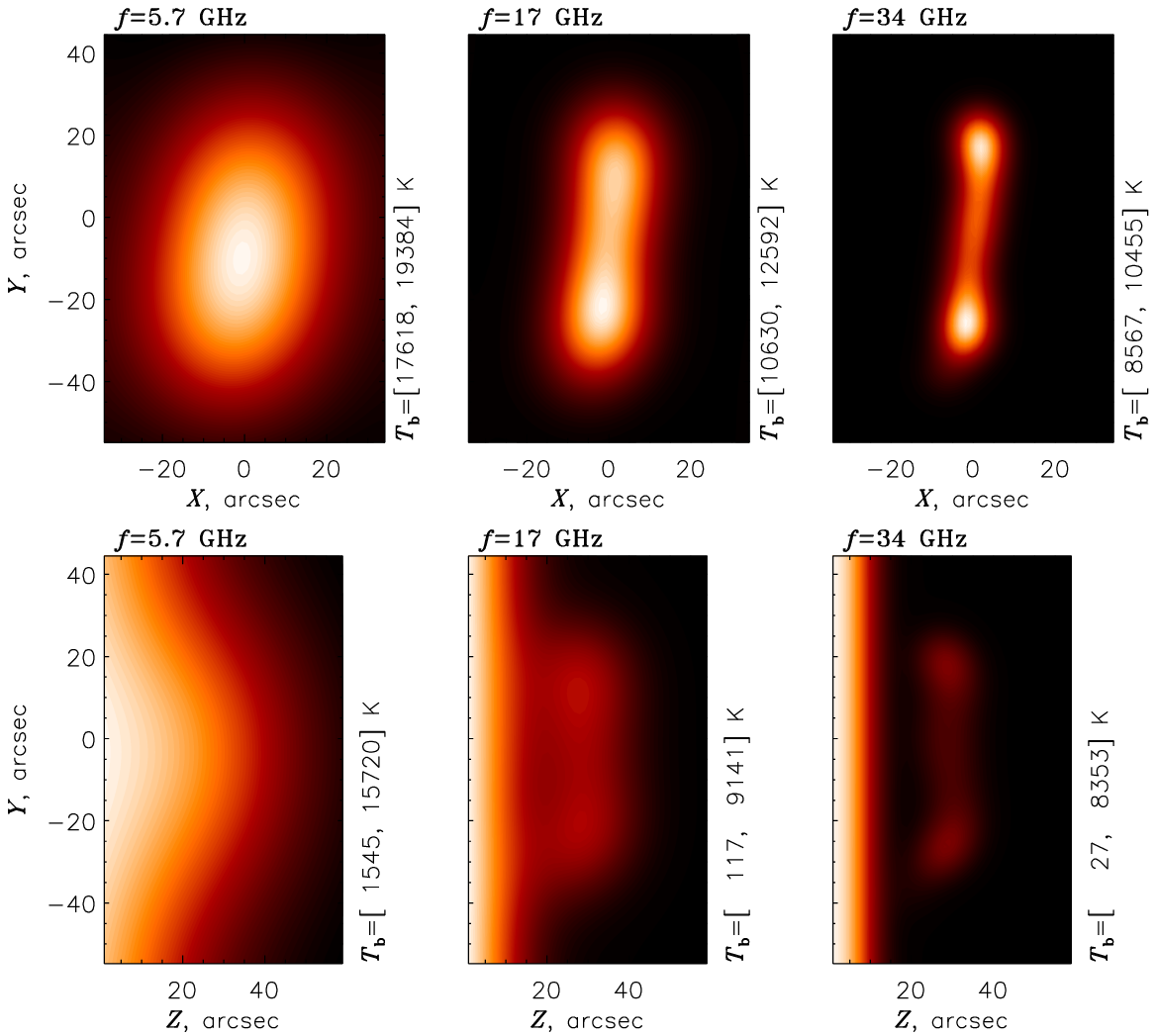}
\caption{Simulated radio images of the filament (same as in Figure \protect\ref{FilamentFinal}) convolved with the instrumental response functions at different frequencies. At the right side of each panel, the range of brightness temperatures in the image is given.}
\label{FilamentFinalSM}
\end{figure}

\citet{xia14} reported the formation of a simulated filament by in situ condensation in a pre-existing flux rope using thermodynamic MHD simulations. The final stable state is shown in Figure \ref{filamentxc}. We have calculated the radio emission from this model filament, and Figures \ref{FilamentFinal}--\ref{FilamentFinalSM} demonstrate the simulated radio images at three frequencies; the final stage of the filament evolution is shown. The plasma density in the filament reaches $10^{10}$ $\textrm{cm}^{-3}$ and the temperature can drop to about $20\,000$ K. The magnetic field in the simulation box does not exceed 11 G; therefore the contribution of the gyroresonance emission is negligible. The chosen frequencies correspond to the working frequencies of the Siberian Solar Radio Telescope (5.7 GHz) and the Nobeyama Radioheliograph (17 and 34 GHz).

Figure \ref{FilamentFinal} shows the non-smoothed images. In addition, to account for a limited resolution of the existing radio telescopes, the calculated emission maps were convolved with the model (symmetric Gaussian-shaped) instrumental response functions. For the filaments at the limb, the image smoothing procedure takes into account the contribution of the solar disk (with the typical intensity of the quiet Sun) at the corresponding (left) boundary of the image. The beam widths are taken to be 21'', 10'' and 5'' at the frequencies of 5.7, 17 and 34 GHz, respectively. The resulting (smoothed) radio maps are shown in Figure \ref{FilamentFinalSM}.

Since the model filament is hotter than real ones ($T>20\,000$ K), the filament on the disk looks brighter than the underlying chromosphere. In addition, the estimation of $T\sim 20\,000$ K refers only to the central part of the filament, while the plasma in the transition region between the filament and the surrounding corona is much hotter and still has a sufficiently high density. As a result, this transition region makes a dominant contribution to the radio emission, which is seen especially well in the non-smoothed image at 5.7 GHz in Figure \ref{FilamentFinal}; the brightness temperature reaches 49\,000 K. Smoothing due to the limited spatial resolution removes this fine filament structure and lowers the image contrast (see Figure \ref{FilamentFinalSM}). In general, the simulated radio images at the disk center look qualitatively different from the observed ones due to the higher filament temperature. Nevertheless, if the filament is observed against a bright active region \citep[like, e.g., in the event reported by][]{gop13}, it will appear as a darker stripe.

In contrast, the filament above the limb looks qualitatively similar to the observed ones, i.e., like a bright prominence (although the brightness temperatures are higher than in real observations). At the frequencies of $\lesssim 10$ GHz, the filament is optically thick and its brightness temperature (in the non-smoothed images) reflects an effective temperature of the transition region between the filament and corona ($\sim 42\,000$ K at 5.7 GHz). At higher frequencies, the filament becomes optically thin. From the non-smoothed images at 17 and 34 GHz we can conclude that its brightness temperature decreases with frequency $f$ approximately as $T_{\mathrm{b}}\propto f^{-2}$; this behaviour is typical of the optically thin free-free emission and has been detected in the multi-frequency observations of solar filaments. The visible structure of the filament at 17 and 34 GHz (again, in non-smoothed images) reflects the plasma and temperature distribution. Since the considered model filament is relatively small, convolving the simulated images with the instrument response functions removes any fine structure (see Figure \ref{FilamentFinalSM}); at lower frequencies, the filament merges with the solar disk.

\subsection{Comparison of the models}
As has been said above, the parameters of our MHD models have been chosen to illustrate different regimes of the flux rope formation. Besides the different magnetic field strengths, these models can have (due to the numerical restrictions) different simulation box sizes and different initial plasma parameters. Therefore a direct quantitative comparison of the models is hardly possible. Nevertheless, we summarize here some representative quantities in order to provide a more comprehensive picture of our simulations and to analyze general trends.

\begin{figure}
\centerline{\includegraphics{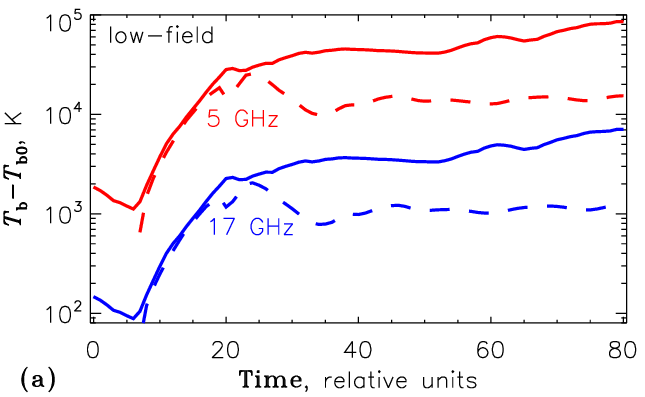}\includegraphics{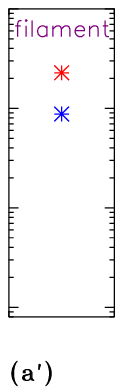}}
\centerline{\includegraphics{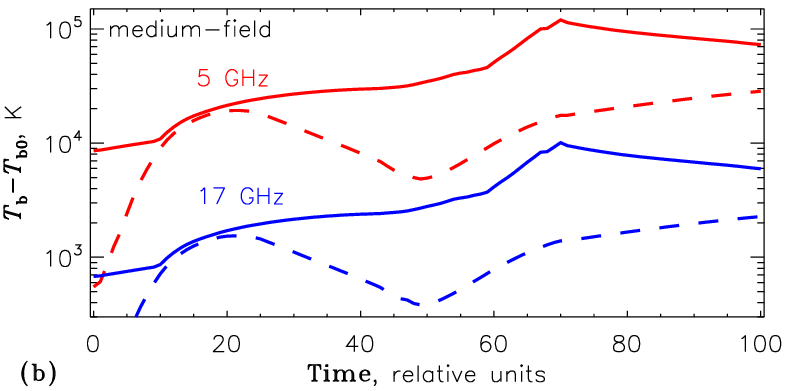}}
\centerline{\includegraphics{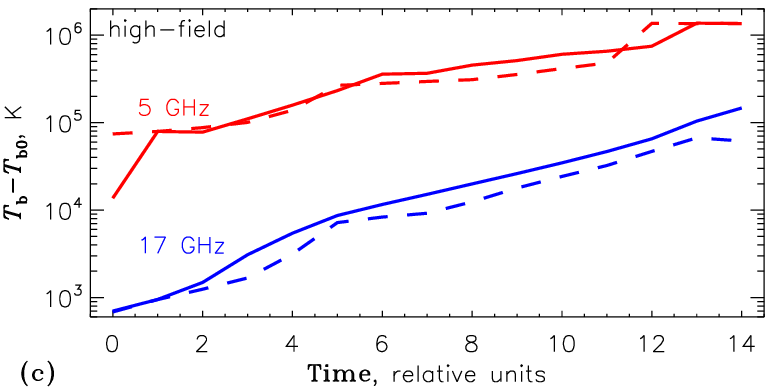}}
\caption{Temporal evolution of the radio emission from the simulated flux ropes (located at the center of the solar disk). The relative radio brightening is shown, where the background level $T_{\mathrm{b0}}$ corresponds to an unperturbed unmagnetized corona. a) The quiet-Sun filament channel (low-field model); a$'$) the simulated filament (no time dependence); b) the medium-field active-region filament channel model; c) the high-field active-region filament channel model. Solid lines: emission from a flux rope footpoint; dashed lines: emission from the flux rope center. In panel (a$'$), the asterisks ($\ast$) show the emission from the center of the simulated filament in its final stable state.}
\label{LightCurves}
\end{figure}

Figure \ref{LightCurves} demonstrates the temporal evolution of the flux ropes. Namely, we plot here the microwave light curves at two frequencies (5 and 17 GHz) for two chosen locations (pixels) in the image plane; the flux ropes are assumed to be located at the center of the solar disk. The flux rope center corresponds to the coordinates $x=0$, $y=0$ in Figures \ref{QuietEvol}--\ref{FilamentFinalSM} (top views), while the tentative flux rope footpoint location at each time is defined as the point of the brightest optically thin radio emission. The light curves for three isothermal models are shown; in the top row of Figure \ref{LightCurves} we show also the corresponding brightness temperatures for the non-isothermal filament model (the endstate and the central pixel only). The background (initial) emission is subtracted, i.e., the values in Figure \ref{LightCurves} represent the relative microwave brightening caused by the plasma and magnetic field concentration.

Evidently, different models have different evolution timescales: in the models with stronger magnetic field, the flux ropes are developing faster. As has been said above, the microwave emission in the low- and medium-field models is produced due to the free-free mechanism in the optically thin regime; as a result, the 5 and 17 GHz light curves mimic each other (they differ by a constant factor). The emissions from the flux rope center and from the footpoints demonstrate different dynamics; at later stages of the evolution, the emission from the footpoints region becomes considerably brighter due to larger projected column density (for the considered orientation). In contrast, in the high-field model at low frequencies the gyroresonance emission mechanism becomes increasingly dominant with time; at later stages of the evolution, the emission brightness temperature at $\lesssim 5$ GHz approaches the plasma temperature (both at the flux rope center and at the footpoints). The intensity of the optically thin free-free emission (e.g., at 17 GHz) in this model steadily increases with time, with a relatively low contrast between the flux rope center and the footpoints. Finally, the emision brightness temperatures for the filament core are higher than the corresponding values for the isothermal low-field flux rope, due to both the higher plasma density and the lower temperature in the filament.

\begin{figure}
\includegraphics{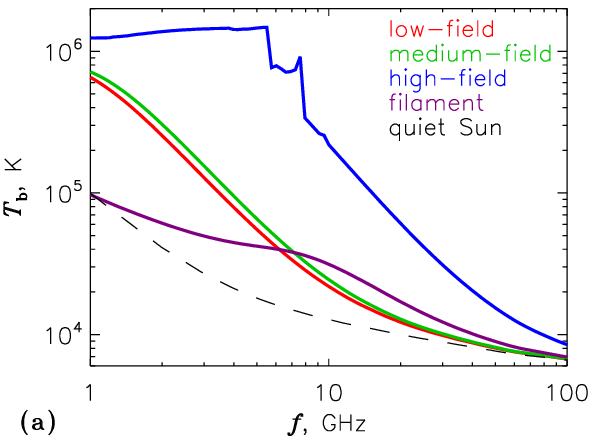}%
\includegraphics{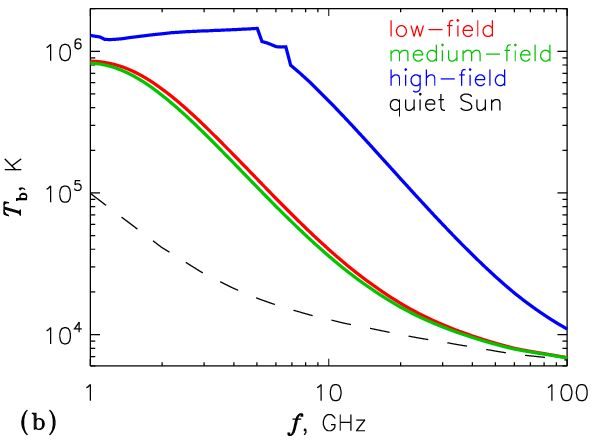}
\caption{\bf Spectra of the radio emission from the simulated flux ropes (located at the center of the solar disk) in their endstates. a) Emission from the flux rope center (emission from the center of the simulated filament is also shown); b) emission from a flux rope footpoint.}
\label{SpectraFinalState}
\end{figure}

Figure \ref{SpectraFinalState} demonstrates several examples of broadband microwave spectra of the simulated flux ropes and the filament (in their endstates). Like in the previous Figure, the flux ropes are assumed to be located at the center of the solar disk; the emissions from the flux rope center and footpoint are considered. The typical spectrum of the quiet Sun \citep{lan08} is plotted for reference. We can see that the spectra for the low- and high-field models almost coincide; they look as the typical isothermal free-free spectra (although the plasma densities can be somewhat different in different models and parts of the flux rope). In the high-field model, the spectra include the optically thick gyroresonance part corresponding to the third gyrolayer (at the frequencies of up to $5-6$ GHz), the optically thin gyroresonance part corresponding to the fourth gyrolayer (up to $7-8$ GHz) and the optically thin free-free part (at higher frequencies); such spectral structure can be potentially used for diagnosing the coronal magnetic field, provided that we have observations with sufficiently high spectral and spatial resolutions. Finally, the filament spectrum reflects the dominant emission mechanism in that model, i.e., the free-free emission in a multi-thermal source.

\section{Conclusions and outlook}\label{conclusion}
We have demonstrated the ability to generate virtual radio views on solar flux ropes from MHD simulations, where full thermodynamic information is available. They can be contrasted with the actual radio views at different frequencies, and may serve to verify inversions (i.e., estimates of the field strengths and density-tem\-pe\-ra\-tu\-re conditions) made from the radio data. We have shown how the views for different orientations (either on the disk or at the limb) can be produced, with (possibly) taking into account the instrument-specific viewing conditions.

The flux ropes with low to medium magnetic field strengths produce radio emission due to the thermal bremsstrahlung process; a deformation from an arcade to a flux rope configuration is seen as a clear change from a photospheric-inversion-line-aligned emission feature to a footpoint-dominated double-spot pattern. At higher field strengths, the gyroresonance emission mechanism dominates at $\lesssim 7$ GHz; the resulting radio views can be associated with the frequently observed neutral-line-associated microwave sources. The medium-field model representing the intermediate case (with the field strength of the order of 100 G) provides the results qualitatively similar to the low-field model; nevertheless, it is an important illustration of the transition from the low-field to high-field MHD models.

For the virtual filament studied, our on-disk images show emission instead of the expected brightness depression. In the limb views, the filament can clearly be seen above the limb, especially at higher frequencies ($\gtrsim 17$ GHz). We conclude that the approach to simulating the filaments seems correct, although the simulation parameters (i.e., the filament temperature and dimensions) still differ from the typical parameters of the observed filaments. 

Since only a few MHD models have been analyzed (because, as said above, the MHD simulations are very expensive computationally), our results are by no means comprehensive. Instead, this study should be considered as a first attempt to combine the MHD simulations of the flux ropes and filaments with the 3D simulations of the radio emission. The results obtained, although somewhat qualitative, demonstrate how the flux rope evolution affects the radio emission; they show the directions for further, more advanced and realistic simulations.

\begin{acks}
This work was supported in part by the Russian Foundation of Basic Research (grants 14-02-91157, 15-02-03717 and 15-32-51171) and by a Marie Curie International Research Staff Exchange Scheme ``Radiosun'' (PEOPLE-2011-IRSES-295272). RK and CX acknowledge support from the project GOA/2015/014 (KU Leuven), the Interuniversity Attraction Poles Programme initiated by the Belgian Science Policy Office (IAP P7/08 CHARM), and the FWO. Part of the simulations used the VSC (Flemish Supercomputer Center) funded by the Hercules foundation and the Flemish Government.
\end{acks}

\end{article}
\end{document}